\documentclass[aps,10pt,prd,reprint,showpacs,superscriptaddress]{revtex4-1}
\usepackage{booktabs,amsmath}
\usepackage{amssymb}
\usepackage{array}
\usepackage{mathtools}
\usepackage{amsthm}
\usepackage{newlfont}
\usepackage{esdiff}
\usepackage{multirow}
\usepackage{graphics}
\usepackage{graphicx}
\usepackage{ae,aecompl,color}
\usepackage{tabularx}
\usepackage{bm}
\usepackage[bookmarks=true,bookmarksopen=true,bookmarksnumbered=true,bookmarksopenlevel=3]{hyperref}

\begin{document}
\newcommand{\bd}{\begin{document}}
\newcommand{\ed}{\end{document}}
\newcommand{\bc}{\begin{center}}
\newcommand{\ec}{\end{center}}
\newcommand{\bfr}{\begin{flushright}}
\newcommand{\efr}{\end{flushright}}
\newcommand{\lt}{\left}
\newcommand{\rt}{\right}
\newcommand{\vs}{\vspace}
\newcommand{\hs}{\hspace}
\newcommand{\beq}{\begin{equation}}
\newcommand{\eeq}{\end{equation}}
\newcommand{\lb}{\linebreak}
\newcommand{\pb}{\pagebreak}
\newcommand{\mb}{\makebox}
\newcommand{\fb}{\framebox}
\newcommand{\mc}{\multicolumn}
\newcommand{\ben}{\begin{enumerate}}
\newcommand{\een}{\end{enumerate}}
\newcommand{\bit}{\begin{itemize}}
\newcommand{\eit}{\end{itemize}}
\newcommand{\un}{\underline}
\newcommand{\lefq}{\lefteqn}
\newcommand{\ba}{\begin{array}}
\newcommand{\ea}{\end{array}}
\newcommand{\beqa}{\begin{eqnarray}}
\newcommand{\eeqa}{\end{eqnarray}}
\newcommand{\beqas}{\begin{eqnarray*}}
\newcommand{\eeqas}{\end{eqnarray*}}
\newcommand{\bfg}{\begin{figure}}
\newcommand{\efg}{\end{figure}}
\newcommand{\bds}{\begin{displaymath}}
\newcommand{\eds}{\end{displaymath}}
\newcommand{\btb}{\begin{tabbing}}
\newcommand{\etb}{\end{tabbing}}
\newcommand{\para}{\parallel}
\newcommand{\pad}{\partial}
\newcommand{\nn}{\nonumber}
\newcommand{\la}{\leftarrow}
\newcommand{\ra}{\rightarrow}
\newcommand{\lgla}{\longleftarrow}
\newcommand{\lgra}{\longrightarrow}
\newcommand{\La}{\Leftarrow}\newcommand{\Ra}{\Rightarrow}
\newcommand{\Lra}{\Leftrightarrow}
\newcommand{\Lgla}{\Longleftarrow}
\newcommand{\Lgra}{\Longrightarrow}
\newcommand{\lan}{\langle}
\newcommand{\ran}{\rangle}
\renewcommand{\a}{\alpha}
\renewcommand{\b}{\beta}
\newcommand{\g}{\gamma}
\newcommand{\G}{\Gamma}
\renewcommand{\d}{\delta}
\newcommand{\eps}{\epsilon}
\newcommand{\Th}{\Theta}
\newcommand{\s}{\sigma}
\newcommand{\lam}{\lambda}
\newcommand{\D}{\Delta}
\newcommand{\vare}{\varepsilon}
\newcommand{\pr}{\prime}
\newcommand{\ro}{\rho}
\newcommand{\nab}{\nabla}
\newcommand{\m}{\mu}
\newcommand{\n}{\nu}
\newcommand{\Sg}{\Sigma}
\newcommand{\p}{\pi}
\newcommand{\R}{I\!\!R}
\newcommand{\om}{\omega}
\newcommand{\Om}{\Omega}
\newcommand{\ze}{\zeta}
\newcommand{\vart}{\vartheta}
\newcommand{\tri}{\triangle}
\newcommand{\f}{\frac}
\newcommand{\iny}{\infty}
\newcommand{\pro}{\propto}
\renewcommand{\arraystretch}{1.25}
\title{Exact solutions of the (2+1) Dimensional Dirac equation in a constant magnetic field in the presence of a minimal length } 
%
%
\author{\textsc{L.~Menculini}}
\affiliation{Dipartimento di Fisica, Universit\`a degli Studi di Perugia, Via A.~Pascoli, I-06123 Perugia, Italy}
\author{\textsc{O.~Panella}}
\affiliation{Istituto Nazionale di Fisica Nucleare, Sezione di Perugia, Via A.~Pascoli, I-06123 Perugia, Italy}
\email[({\bf Corresponding Author}) Email: ]{orlando.panella@pg.infn.it }

\author{\textsc{P.~Roy}}
\affiliation{Physics and Applied Mathematics Unit, Indian Statistical Institute, Kolkata, India}

\date{\today}

\begin{abstract}
We study the (2+1) dimensional Dirac equation in an homogeneous  magnetic field  (relativistic Landau problem) within  a minimal length, or generalized uncertainty principle -GUP-,  scenario. We derive exact solutions for a given  explicit representation of the GUP and provide  expressions of the wave functions in the momentum representation. We find that in the minimal length case the degeneracy of the states is modified and that there are states that do not exist in the ordinary quantum mechanics limit ($\beta \to 0$). We also discuss the massless case which may find application in describing the behavior of charged fermions in new materials like  Graphene.
\end{abstract}

\pacs{03.65.Pm,03.65.Ge,12.90.+b,02.40.Gh}
\maketitle

\section{Introduction}
In recent years there has been extensive research on the minimal length formalism. The concept of a minimal length has emerged from various studies on quantum gravity \cite{garay},
perturbative string theory \cite{gross} and  black holes \cite{magg}. See~\cite{Hossenfelder:2012jw} for a recent review. A consequence of the presence of a minimal length is that the Heisenberg uncertainty relation becomes modified and this results in UV/IR mixing. Consequently it is meaningful to study quantum mechanics in the presence of a minimal length \cite{kempf1,kempf2,kempf3,kempf4}. In particular, exact solutions
of various non relativistic~\cite{chang,dadic,gemba,brau,fityo,akhoury,benczik} and relativistic problems~\cite{quesne1,nouicer,quesne2,jana} have been obtained in the presence of a minimal length ($\Delta x_0=\hbar\sqrt{\beta}$).  Approaches have also been discussed that try to  incorporate a minimal length in the quantum field theory formalism and explicit calculations of the Casimir effect~\cite{Frassino:2011aa,Dorsch:2011qf} and the Casimir-Polder interactions~\cite{Panella:2007kd}  within a generalized uncertainty principle have been derived. 

{\color{black}
Many of the studies available in the literature deal with specific calculations, report on the regularizing properties of the minimal length, and also have the purpose of deriving upper bounds on the minimal length via comparison with experimental measurements, where possible.   
The authors of ref.~\cite{Bouaziz:2007gs,Bouaziz:2010hc} for instance solve the inverse square potential exactly in arbitrary dimensions and show how the minimal length acts as a natural cut-off regulator. 
In ref.~\cite{Stetsko:2007ze} the authors study the scattering problem within a GUP scenario and derive the dependence on $\hbar\sqrt{\beta}$ of the scattering amplitude and cross section. We may note that upper bounds of quite different magnitude have been derived. In ref.~\cite{Benczik:2002tt} the semiclassical limit of the GUP scenario has been addressed and a quite impressive constraint on the minimal length has been derived by computing the perihelion shift in a central force potential. Comparing it with the observed precession of the perihelion of Mercury results in $(\Delta x)_{min}=\hbar\sqrt{\beta} < 10^{-68}\ $m, some 33 orders magnitude below the Planck length ($L_P=\sqrt{\frac{\hbar G}{c^3}} = 1.16 \times 10^{33}\ $m). 
Other interesting constraints come from including the corrections due to the minimal length to the hydrogen atom spectrum and computing  the Lamb shift. The accurate measurements available for the Lamb shift allow to derive and upper bound on the minimal length of the order of the electroweak scale:
$(\Delta x)_{min}= \hbar\sqrt{\beta} < 10^{-17}\ $m
\cite{Benczik:2005bh,Stetsko:2006mm}. See also the recent works in~\cite{Das:2008kaa} and \cite{ Ali:2011fa} for further discussions about the minimal length phenomenology using a somewhat different GUP representation than the one taken up here.
}

Here we propose to study a relativistic quantum mechanical problem, namely, the $(2+1)$ dimensional Dirac equation in the presence of a minimal length. To be more specific we shall obtain exact solutions {\color{black}(eigenvalues and eigenfunctions)} of the Dirac equation in the presence of a homogeneous magnetic field {\color{black}(relativistic Landau levels --LL--)}. This topic has become quite interesting because of its application to various branches of physics, particularly in condensed matter physics. {\color{black} In passing we may note that due to this growth in the  interest for 2-dimensional electron  systems,  very recently (non relativistic) Landau levels have been for the first time imaged, revealing the expected ring-like internal structure of the wave functions by means of scanning tunneling spectroscopy~\cite {imagingLL}.}  In this context we would like to note that, {\color{black} from the theoretical side}, the Pauli equation has also been studied in the presence of a minimal length \cite{nou}. However, we shall implement the minimal length formalism in the first order Dirac equation rather than after obtaining the second order equations for the spinor components. We shall obtain solutions of the problem after converting the equations for the components into Schr\"odinger like equations with some standard solvable potential. Subsequently the scalar product for the model (which is quite different from the standard one) will be used to determine admissible limits on the angular quantum number $m$ (in the momentum space) and this in turn will be used to determine the spectrum and the corresponding eigenfunctions. A notable feature which emerges from the analysis is that in certain cases the admissible values of the angular quantum number is constrained by a bound which depends on the minimal length. Also, there is a class of states which cease to exist in the limit $\b\ra 0$. Finally it may be noted that apart from being interesting in itself, the massless Dirac equation in $(2+1)$ dimension finds application in condensed matter physics. For example, massless Dirac equation in $(2+1)$ is relevant in describing the motion of electrons in graphene \cite{geim}. In view of this we shall also find the eigenvalue spectrum and the corresponding eigenfunctions in the massless case.
 
{\color{black} 
In addition we shall discuss how our results in the massless case, relevant to graphene, can be used to extract  an upper bound on the minimal length by comparing with experimental measurements of the relativistic Landau levels (LL) in graphene as reported in~\cite{expLL}. Our upper bound on the minimal length derived from comparing  measurements of (electron-electron and electron-hole) transitions  between the first excited Landau levels of graphene from~\cite{expLL} turns out to be $(\Delta x)_{min}=\hbar\sqrt{\beta} < 2.3$ nm 
and is of the same order of magnitude of the bound obtained from considerations of the corrections due to a minimal length on the Casimir effect~\cite{Frassino:2011aa}. In \cite{iontrapping} the authors use the exact solution of the (non relativistic) harmonic oscillator in arbitrary dimensions within a  GUP scenario in order to derive an upper bound on the minimal length referring to measurements on electrons trapped in strong magnetic fields (Penningtrap~\cite{penningtrap}) whose motion is effectively one dimensional. They take advantage of the $n^2$ dependence  of the minimal length correction to the (non relativistic) eigenvalues and derive 
 an upper bound on the minimal length ($\hbar\sqrt{\beta} < 10^{-16}$ m) which is however based on the rather strong assumption of being able to measure the energy eigenvalues up to quite large values of the quantum number ($n \approx 10^8$). Their actual bound ($\hbar \sqrt{\beta } < \frac{15 \times 10^{-9}\,\text{m}}{n}$)  becomes the order of a few nanometers when $n\sim{\cal O}(1)$ and is of the same order of the bound derived in this work (see details in section~\ref{graphene}). 
}

The organization of the paper is as follows: in section \ref{2+1DiracEq} we present the problem and obtain the exact solutions; in section \ref{spectrumWF} we analyze the spectrum and provide explicit expressions for the momentum space wave functions; finally section \ref{discussion} is devoted to a discussion and conclusion.

\section{{(2+1)} dimensional Dirac equation in the presence of a minimal length and within a constant magnetic field}
\label{2+1DiracEq}
In the minimal length formalism the Heisenberg algebra associated with the position coordinates ${\hat x}_i$ and the momentum ${\hat p}_i$ is given by \cite{kempf1,kempf2}:
\beq
[{\hat x}_i,{\hat p}_j]=i\hbar \d_{ij}(1+\b {\bm p}^2)\label{rel1}
\eeq
where $\b>0$ is the minimal length parameter. The corresponding generalized uncertainty principle (GUP) reads:
\beq
\ba{l}
\Delta x_i \Delta p_j\geq \f{\hbar}{2}\d_{ij}[1+\b(\Delta{\mbox{\bf p}})^2+\b\langle{\mbox{\bf p}}\rangle^2]\\
\ea
\eeq
yielding a minimal observable length $\Delta x_0=\hbar\sqrt{\beta}$. A representation of ${\hat x}_i$ and ${\hat p}_i$ which satisfies Eq.(\ref{rel1}) may be taken as
\beq
{\hat x}_i=i\hbar (1+\b {\bm p}^2)\frac{\partial}{\partial p_i},~~~~~{\hat p_i}=p_i\label{rep}
\eeq
from which it also follows that
\begin{align}
\Delta x_i \Delta x_j&\geq \hbar\b|\langle{\hat p}_i{\hat x}_j-{\hat p}_j{\hat x}_i\rangle|\\
\Delta p_i \Delta p_j&\geq 0
\end{align}
It is important to note that the scalar product in this case is not not the usual one but is defined as
\beq
\langle f|g \rangle =\int_{-\infty}^\infty \f{d^2\bm p}{(1+\b {\bm p}^2)}f^*({\bm p})g({\bm p})\label{scalar}
\eeq

Let us now consider the $(2+1)$ dimensional Dirac equation in the presence of a homogeneous magnetic field $\bm{B}= (0,0,B_0) $ with  the corresponding Hamiltonian given by:
\beq
H=c{\bm\s}.({\hat{{\bm p}}+\frac{e}{c}\hat{\bm A}})+\s_z Mc^2
\eeq
where $\bm{\s}=(\s_x,\s_y)$, and $\s_z$ are Pauli matrices and the vector potential is chosen in the symmetric gauge:
\beq
{\hat A}_x=-\f{B_0}{2}{\hat y},~~~~{\hat A}_y=\f{B_0}{2}{\hat x}\, .
\eeq
The eigenvalue problem  reads:
\beq
H\psi=E\psi,~~~~\psi=\left(\ba{cc} \psi^{(1)} \\ \psi^{(2)}\ea\right)\, .
\eeq
Now using the representation (\ref{rep}) the above eigenvalue equation can be written as
\beq H \psi =
\left(\ba{cc}Mc^2 & cP_-\\cP_+ & -Mc^2\ea\right)\left(\ba{c}\psi^{(1)} \\ \psi^{(2)}\ea\right)=E\left(\ba{c}\psi^{(1)} \\ \psi^{(2)}\ea\right) 
\label{H}
\eeq
where we have defined
\beq
P_{\pm}=P_x\pm iP_y=\left(p_x+\frac{e}{c}{\hat A}_x\right)\pm i\left(p_y+\frac{e}{c}{\hat A}_y\right)
\eeq
Written in terms of components, Eq.(\ref{H}) reads 
\beq
P_-\psi^{(2)}=\eps_-\psi^{(1)},~~~~P_+\psi^{(1)}=\eps_+\psi^{(2)},~~~~\eps_{\pm}=\f{E\pm Mc^2}{c}
\label{PpandPm}
\eeq
Then decoupling the components we find

\begin{eqnarray}
P_-P_+\psi^{(1)}&=&\eps^2\psi^{(1)},~~~~P_+P_-\psi^{(2)}=\eps^2\psi^{(2)}, \nonumber \\~~~~\eps^2&=&\eps_+\eps_{-}=\frac{E^2-M^2c^4}{c^2}\label{eigen1}
\end{eqnarray}

Now using the relations (\ref{rep}) we find that
\beq
\ba{lcl}
P_+&=& e^{i\vart}\left[p-\left(1+\b p^2\right)\left(\lambda\partial_p+\frac{i\lambda}{p}\partial_\vart\right) \right] \\
P_-&=& e^{-i\vart}\left[p+\left(1+\b p^2\right)\left(\lambda\partial_p-\frac{i\lambda}{p}\partial_\vart\right) \right]~. \label{polPmETA}
\ea
\eeq
where we have defined
\beq
\label{momentumspace}
\lam= \f{\hbar eB_0}{2c},~~~~p_x=p~\cos\vart,~~~~p_y=p~\sin\vart,~~~~p_x^2+p_y^2=p^2
\eeq


Following~\cite{kempf1} the generator of rotations in the $(x,y)$ plane in the minimal length scenario is defined by:
\begin{equation}
\label{angularmomentum}
\hat{L}_z = \frac{\hat{x}\, \hat{p}_y -\hat{y} \,\hat{p}_x}{1+\beta p^2} = -i\,\hbar\, \partial_\vartheta
\end{equation}
and satisifies the relations $[P_\pm, L_z] =\mp \hbar P_\pm$. It can then be easily verified that the operator $\hat{J}=\hat{L}_z+(\hbar/2)\sigma_z$ commutes with the Hamiltonian in Eq.~\eqref{H}, so that even in the presence of  a minimal length we have a conserved total angular momentum. Note that in the limit $\beta \to 0$ the definition of $L_z$ in Eq.~\eqref{angularmomentum} goes smoothly into the ordinary one.  Thus we see that if $m$ is the quantum number associated to the operator $\hat{L}_z $ the conserved total angular momentum is $j=\hbar (m \pm 1/2)$. Note that although in this instance the angular variable $\vartheta $ is defined  in momentum space, cf Eq.~\eqref{momentumspace}, the quantum number $m$ (associated to the  eigenfunctions $e^{im\vartheta}$ of $\hat{L}_z$) retains its usual meaning of orbital angular momentum quantum number. 

The wave functions may be taken therefore to be eigenstates of the (total) angular momentum (note that the components have to satisfy the intertwining relations in Eq.~\eqref{eigen1}) and we can put them in the form:
\beq
\label{ufun}
\psi^{(1)}_m=u^{(1)}_m(p)e^{im\vart}\, , \qquad \psi^{(2)}_m=u^{(2)}_m(p)e^{i(m+1)\vart}\, . 
\eeq
Then from Eq.(\ref{eigen1}) we obtain:
\begin{multline}
\left\{\vphantom{\frac{a}{b}} p^2+2\lam\left(1+\b p^2\right)\left[m+1-\b\lam\left(p\diff{}{p}-m\right)\right]+\right.\\ \left.-\lam^2\left(1+\b p^2\right)^2\left[\diff[2]{}{p}+\frac{1}{p}\diff{}{p}-\frac{m^2}{p^2}\right]\right\} u^{(1)}_m(p)=\\ \eps^2 u^{(1)}_m(p),\label{u1}
\end{multline}
\begin{multline}
\left\{\vphantom{\frac{a}{b}} p^2+2\lam\left(1+\b p^2\right)\left[m-\b\lam\left(p\diff{}{p}+m+1\right)\right]+\phantom{xx}\right.\\ \left.-\lam^2\left(1+\b p^2\right)^2\left[\diff[2]{}{p}+\frac{1}{p}\diff{}{p}-\frac{(m+1)^2}{p^2}\right]\right\} u^{(2)}_m(p)=\\ \eps^2 u^{(2)}_m(p)~.\label{u2}
\end{multline}
The above equations are still complicated enough to admit direct solutions. However, the solutions may be obtained readily if we can transform the equations to some standard form. To this end we now perform a simultaneous change of wave functions as well as of the variable: 
\beq \label{rx}
\ba{l}
u^{(i)}_m=p^{-\frac{1}{2}}\varphi^{(i)}_m \qquad i=1,2\\
p=\frac{1}{\sqrt{\b}}\tan q,\qquad  q=\frac{x}{2}+\frac{\pi}{4}, \qquad x\in\left[-\frac{\pi}{2},\frac{\pi}{2}\right]
\ea
\eeq
Using the above transformations we obtain from Eq.~\eqref{u1} and Eq.~\eqref{u2}:
\begin{multline} \label{xeqn}
\left\lbrace-\diff[2]{}{x}+\frac{1}{2}\left[\frac{\zeta_i(\zeta_i-1)+\xi_i(\xi_i-1)}{\cos^2(x)}\right]+\right. \\ \left. \frac{1}{2}\Bigl[\xi_i(\xi_i-1)-\zeta_i(\zeta_i-1)\Bigr]\frac{\sin(x)}{\cos^2(x)}\right\rbrace\varphi^{(i)}_m(x)=k^2\varphi^{(i)}_m(x)
\end{multline} 
where
\beq
 k^2=\frac{\eps^2+1/\b}{4\b\lambda^2}~. \label{k}
\eeq
and the parameters $\xi_i$ and $\zeta_i$ are defined as
\begin{alignat}{2}
\zeta_1&=m+\frac{1}{2} & \qquad \xi_1&=m+\frac{3}{2}+\frac{1}{\b\lam}\label{xizeta1}\\
\zeta_2&=m+\frac{3}{2} & \qquad \xi_2&=m+\frac{1}{2}+\frac{1}{\b\lam}\label{xizeta2}
\end{alignat}
One can identify the above Eq.~\eqref{xeqn} as a pair of Schr\"odinger equations (in units where $\hbar^2/(2M) =1$) with the trigonometric Scarf potential of the form:
\beq
\label{potential(x)}
V(x)=\left(\f{\mu^2+\nu^2}{2}-\f{1}{4}\right)\f{1}{\cos^2x}+\f{\mu^2-\nu^2}{2}\f{\sin~x}{\cos^2x}
\eeq
where the parameters $\mu$ and $\nu$ are given in each case ($i=1,2$) in terms of the parameters $\xi_i$ and $\zeta_i$ via:
\begin{equation}
\label{parmunu}
\mu=\xi_i-\frac{1}{2} \, , \qquad  \nu=\zeta_i-\frac{1}{2}\, .
\end{equation}
We may note that the potential $V(x)$ in Eq.~(\ref{potential(x)}) has certain symmetries that will be of use in  writing the solution of our problem.   In particular  $V(x)$ is unchanged by the replacements $\mu, \to -\mu$ and/or $\nu \to -\nu $.
Upon imposing standard boundary conditions on the finite domain $x \in [-\pi/2,+\pi/2]$ or $q\in [0,+\pi/2]$ (normalizability and vanishing of the wave function at the end-points), the  eigenfunctions and eigenvalues of  Eq.~\eqref{xeqn} are readily obtained from~\cite{levai,flugge}:
\begin{eqnarray}\label{levair}
\psi_n(x)&=&C\,\,\displaystyle[z(x)]^{\f{\mu}{2}+\f{1}{4}}[1-z(x)]^{\f{\nu}{2}+\f{1}{4}}\times \nonumber  \\ &&\phantom{xxxxxx}\,\!\!\phantom{F}_2F_1\left(-n,\mu+\nu+1;\nu+1;1-z(x)\right)\nonumber\\
k_n&=&n+\displaystyle\frac{\mu+\nu+1}{2}
\end{eqnarray}
where $z(x)=\displaystyle\f{1-\sin x}{2}=\cos^2(q)$ and $C$ is a normalization constant. Note that $\psi_n(x)$ in Eq.~\eqref{levair} is obtained from~\cite{levai}, via the substitution: $\mu \leftrightarrow \nu , z \to 1-z $ which is easily verified to be a symmetry of the potential $V(x)$ in Eq.~\eqref{potential(x)}.
\begin{table*}[t!]
\begin{tabularx}{\textwidth}{ c | >{\centering\arraybackslash}X | c } \hline \hline
$m$ & ${ \varphi^{(1)}_{n,m}}$ & ${ k^2}$  \\ 
\hline 
${m}\geq 0$ & $\left(\sin q\right)^{\zeta_1}\left(\cos q\right)^{\xi_1}\prescript{}{2}F_1\left(-n,n+\zeta_1+\xi_1,\zeta_1+\frac{1}{2};\sin^2q\right)$ & $\frac{1}{4}\left(2n+\zeta_1+\xi_1\right)^2$  \\
\hline  
$-\frac{3}{2}-\frac{1}{\lam\b}<{ m}\leq -1$ & $\left(\sin q\right)^{1-\zeta_1}\left(\cos q\right)^{\xi_1}\prescript{}{2}F_1\left(-n,n+1-\zeta_1+\xi_1,\frac{3}{2}-\zeta_1;\sin^2q\right)$& $\frac{1}{4}\left(2n+1-\zeta_1+\xi_1\right)^2$ \\
\hline
${ m}<-\frac{1}{2}-\frac{1}{\lam\b}$ & $\left(\sin q\right)^{1-\zeta_1}\left(\cos q\right)^{1-\xi_1}\prescript{}{2}F_1\left(-n,n+2-\zeta_1-\xi_1,\frac{3}{2}-\zeta_1;\sin^2q\right)$& $\frac{1}{4}\left(2n+2-\zeta_1-\xi_1\right)^2$ \\ 
\hline \hline
\end{tabularx}
\caption{$\varphi^{(1)}_{n,m}$ and the corresponding energy values for different values of $m$. In this case a solution to Eq.~\eqref{parmunu} is $\mu=m+1+\frac{1}{\beta\lambda}$ and $\nu=m$. The range $m\ge 0$ is obtained solving in terms of $m$ the constraints: $\mu, \nu >-\frac{1}{2}$ (or equivalently $ \zeta_1, \xi_1>0)$.  The range $ -\frac{3}{2}-\frac{1}{\beta\lambda}<m\le -1$ is obtained solving  the constraints: $\mu > -\frac{1}{2}, \nu <\frac{1}{2}$ (or equivalently $ (1-\zeta_1)>0, \xi_1>0)$. The range $  m< -\frac{1}{2}-\frac{1}{\beta\lambda}$ is obtained solving  the constraints: $\mu , \nu <\frac{1}{2}$ (or equivalently $ (1-\zeta_1)>0, (1-\xi_1)>0)$. The fourth possible constraint $\mu<\frac{1}{2}, \nu>-\frac{1}{2}$ does not admit solutions for any value of $m$.} \label{tab:up}
\end{table*}
\begin{table*}[t!]
\begin{tabularx}{\textwidth}{ c | >{\centering\arraybackslash}X | c } \hline \hline 
$m$ & ${\varphi^{(2)}_{n,m}}$ & ${ k^2}$ \\ 
\hline 
${ m}\geq 0$ & $\left(\sin q\right)^{\zeta_2}\left(\cos q\right)^{\xi_2}\prescript{}{2}F_1\left(-n,n+\zeta_2+\xi_2,\zeta_2+\frac{1}{2};\sin^2q\right)$ & $\frac{1}{4}\left(2n+\zeta_2+\xi_2\right)^2$  \\
\hline  
$-\frac{1}{2}-\frac{1}{\lam\b}<{ m}\leq -1$ & $\left(\sin q\right)^{1-\zeta_2}\left(\cos q\right)^{\xi_2}\prescript{}{2}F_1\left(-n,n+1-\zeta_2+\xi_2,\frac{3}{2}-\zeta_2;\sin^2q\right)$& $\frac{1}{4}\left(2n+1-\zeta_2+\xi_2\right)^2$ \\
\hline
${ m}<\frac{1}{2}-\frac{1}{\lam\b}$ & $\left(\sin q\right)^{1-\zeta_2}\left(\cos q\right)^{1-\xi_2}\prescript{}{2}F_1\left(-n,n+2-\zeta_2-\xi_2,\frac{3}{2}-\zeta_2;\sin^2q\right)$& $\frac{1}{4}\left(2n+2-\zeta_2-\xi_2\right)^2$ \\ \hline \hline
\end{tabularx}
\caption{$\varphi^{(2)}_{n,m}$ and the corresponding energy values for different values of $m$. In this case a solution to Eq.~\eqref{parmunu} is $\mu=m+\frac{1}{\beta\lambda}$ and $\nu=m+1$. The three ranges of the $m$ values are found solving the same constraints described in the caption of Table~\ref{tab:up}.  Note that in this case the fourth constraint $\mu < \frac{1}{2}$, $\nu > -\frac{1}{2}$ has solutions for $m$ in the range $ -\frac{3}{2} <m < \frac{1}{2}- \frac{1}{\beta\lambda}$ which is meaningful only for $\frac{1}{\beta\lambda} < 2$. However the minimal length is physically expected to be a small quantity and  we have indeed  $\frac{1}{\beta\lambda} >>1$. See the discussion in the text and Eq.~\eqref{constr}. So this possibility will be ignored throughout.} \label{tab:down}
\end{table*}

The vanishing of the wave-function at the end-points (i.e. $q=0$ and $q=\pi/2$) is ensured by enforcing the following constraints: 
(a) $\mu>-1/2$ and $\nu > -1/2$; (b) $\mu<1/2$ and $\nu < 1/2$; (c) $\mu >  -1/2 \text{\ and\ }\nu< 1/2 $; (d) $\mu <  1/2 \text{\ and\ }\nu>- 1/2 $.
Solving the parameters $\mu ,\nu$ in Eqs.~(\ref{parmunu}) in terms of the angular momentum quantum number $m$ provides with the three ranges (of $m$) in Tables \ref{tab:up} and \ref{tab:down}. Note that one of the constraints does not have solution for any value of $m$. The wave-functions in Tables \ref{tab:up} and \ref{tab:down} are then obtained from Eq.~\eqref{levair}. In the second and third row of both tables repeated use is made of  the fact that the potentials   in Eq.~\eqref{xeqn} are invariant under the reparametrization:
\beq
\zeta_i\ra 1-\zeta_i\,, \qquad \text{and/or} \qquad \xi_i \ra 1-\xi_i\,. \label{replace}
\eeq

We conclude this section with a final important remark. While we have applied  standard boundary conditions in the finite $x$ (or $q$) domain (normalizability and vanishing of the wave function at the end-points) it is interesting to note that these can be transported back to the physical (radial) $p$-space  of our original Dirac problem and can be given a physically sound interpretation.

The normalization integral of the Dirac spinor reads:
\begin{equation}
\langle \psi | \psi \rangle  = \int  \frac{d^2 \bm{p}}{1+\beta \bm{p}^2}
\left[\left(\psi^{(1)}\right)^*\psi^{(1)}+\left(\psi^{(2)}\right)^*\psi^{(2)}\right]
\end{equation}
and normalizability of the spinor solution is guaranteed, if both radial components components satisfy:
\begin{equation}
\int_0^\infty \frac{pdp}{1+\beta p^2}\,  |u(p)|^2\quad  < \quad \infty
\end{equation}
Because of the deformation of the measure, introduced by the minimal length, the asymptotic behavior of the $u^{(i)}(p)$ functions that ensures  such condition is:
\begin{equation} 
u^{(i)}(p)\quad {\sim \atop {p\to \infty}} \quad \frac{1}{p^\chi}   \quad \chi>0
\end{equation}
In the \emph{reduced} problem, where $\varphi^{(i)}(p)= p^{1/2}\, u^{(i)}(p)$,
unless $\chi$ is large enough ($\chi>1/2$) the wave function $\varphi^{(i)}(p)$ \emph{will not vanish} as $p\to \infty$.
Thus in this sense we conclude that normalizability alone of the $u^{(i)}(p)$ wave functions does not warrant that the reduced wave-functions $\varphi^{(i)}(p)$  vanish at $p\to \infty$ (or in $q$-space at $q=\pi/2$), while $\left.\varphi^{(i)}(q)\right|_{q=\frac{\pi}{2}}=0$ is the standard boundary condition which we have implemented in building up the results of Tables~\ref{tab:up} and \ref{tab:down}. Note that for a vanishing minimal length ($\beta \to 0$) the measure reduces to the standard one and the normalizability of the $u^{(i)}(p)$ requires instead $\chi>1$ which would ensure that  the reduced wave function  vanishes as $p\to \infty$ (or $q=\pi/2$). 

On the other hand in our relativistic Dirac problem the energy integral computed from the quantum Dirac hamiltonian in Eq.~\eqref{H} is:\begin{eqnarray}
\label{energy_integral}
&&\langle \psi | H | \psi \rangle  =\int  \frac{d^2 \bm{p}}{1+\beta \bm{p}^2}
\left[ Mc^2\,\left(\psi^{(1)}\right)^*\psi^{(1)}+\right.\nonumber \\&&\phantom{xxxxxxxxxx}\left.  \left(\psi^{(1)}\right)^*\, {cP_-}\,\psi^{(2)}
-Mc^2\,\left(\psi^{(2)}\right)^*\psi^{(2)}+\right. \nonumber \\
&&\left.\phantom{xxxxxxxxxxxx}\left(\psi^{(2)}\right)^*\, {cP_+}\,\psi^{(1)}\right]
\end{eqnarray}
Now require \emph{in addition} the finiteness of the energy integral. 
This means that the second and fourth integrals in Eq.~\eqref{energy_integral} must be finite. The operators $P_{\pm}$ as given in Eq.~\eqref{polPmETA} are linear in the radial momentum $p$ (as it is expected from a Dirac Hamiltonian).
Then assuming that as $p \to \infty$ the components behave as $u^{(i)} \sim p^{-\chi}$ (with $\chi>0$ to ensure normalizability) the asymptotic beahvior of the integrands in the second and fourth integral in Eq.~\eqref{energy_integral} is:
\begin{equation}
\frac{p}{1+\beta p^2} \,\frac{1}{p^\chi} \, {\cal O}(p)\, \frac{1}{p^\chi}\qquad  {\sim\atop p \to \infty}\qquad \frac{1}{p^{2\chi}}
\end{equation}
which turn out to be integrable only if $2\chi>1$ or  $\chi>1/2$ which, for example in the case of the first row of Table~\ref{tab:up}   ($\chi=\xi_1+1/2$), translates into $\xi_1 >0$ or $\mu>-1/2$ ensuring that $\left.\varphi^{(1)}\right|_{q=\frac{\pi}{2}} =0 $.

Similar considerations can be performed as regards  the behavior of the wave-functions at $p=0$, and again the vanishing of the $\left.\varphi^{(i)}(q)\right|_{q=0}$  is ensured by the finiteness of the energy integral. Note that in the limit $p\to 0$  the operators $P_\pm$ in Eq.~\ref{polPmETA} will be dominated by the derivative terms ($\partial_p$) which give an extra inverse power of $p$ in the third and fourth integrals of Eq.~\ref{energy_integral}. In particular we have verified that all the conditions discussed to deduce  Tables~\ref{tab:up} and \ref{tab:down} can be deduced in the radial $p$-space by imposing the finiteness of the energy integral.

Let us conclude these considerations with a final observation. 
Physically we would expect that the wave functions $u^{(i)}(p)$ and thus the $\psi^{(i)}(p)$ have always a  regular behavior at $p=0$.  However  this is not excluded by the  boundary condition that we have imposed in the reduced problem.
From Eq.~\eqref {wfp} we see that the condition to require that the $u^{(i)}$ function would not be divergent at $p=0$ is $\zeta_i -1/2 \ge 0$  while the condition that we have imposed is the less restrictive one $\zeta_i >0 $.
We notice however in Tables \ref{tab:positive}, \ref{tab:1negative} and \ref{tab:2negative} that our wave functions never diverge for $p \to 0 $. This can be understood by the fact that the $\zeta_i $ of our problem are not continuous parameters but are instead discrete because  they depend on the orbital angular momentum quantum number $m$.
Indeed in the case of Eq. ~\ref{wfp} $\zeta=m+1/2$ and the condition $\zeta>0$ reduces to $m>-1/2 $ which is effectively equivalent to $m\ge 0$ (since $m$ is integer) or $\zeta-1/2\ge 0$.


\begin{table*}[t!]
\caption{\label{tab:positive}Energy levels and the corresponding wave functions for $m\geq 0$. A given energy level with $n+m=N$ has a finite degeneracy $D=N+1$.} 
\begin{ruledtabular}
\begin{tabular}{l} 
$E_{n,m}=\sqrt{M^2c^4+2\hbar eB_0 c \left(n+m\right)\left[1+\b\frac{\hbar eB_0}{2c}\left(n+m\right)\right]} $\ \ \ \ 
 $ n=0,1,2,\ldots$\ \ \ \
$\psi_{n,m} =
\begin{pmatrix} 
\psi_{n,m}^{1} \\
\psi_{n,m}^{2} 
\end{pmatrix} $\\  \hline 
$ \psi_{n,m}^{(1)}=C_1\frac{p^{m}\prescript{}{2}F_1\left(-n,n+2(m+1)+\frac{1}{\lambda\b},m+1,\frac{\b p^2}{1+\b p^2}\right)}{\left(1+\b p^2\right)^{m+1+\frac{1}{2\lambda\b}}}e^{im\vart}$ \, \, \, \, 
$ \psi^{(2)}_{n,m}=C_2\frac{p^{(m+1)}\prescript{}{2}F_1\left(-n,n+2(m+1)+\frac{1}{\lambda\b},m+2,\frac{\b p^2}{1+\b p^2}\right)}{\left(1+\b p^2\right)^{m+1+\frac{1}{2\lambda\b}}}e^{i(m+1)\vart}$ \\
\end{tabular} 
\end{ruledtabular}
\end{table*}
\begin{table*}[t!]
\caption{Energy levels and the corresponding wave functions for $-\frac{1}{2}-\frac{1}{\lam\b}<m\leq -1$. The degeneracy $D$ of these levels is finite and explicitly given by $D= [\frac{1}{2}+\frac{1}{\lambda \beta}]$. In the limit of a vanishing minimal length ($\beta \to 0 $) the degeneracy of these energy levels becomes infinite ($D\to\infty$).}\label{tab:1negative}
\begin{ruledtabular}
\begin{tabular}{lr}
$E_{0}=Mc^2$\ \ \ \  $\psi_{0,m}=\begin{pmatrix}
0 \\
\psi_{0,m}^{(2)} 
\end{pmatrix}$ &  $\begin{matrix} E_{n}=\sqrt{M^2c^4+2\hbar eB_0 c n\left(1+\b\frac{\hbar eB_0}{2c}n\right)} \\ n=1,2,\ldots \end{matrix}$ 
\  $\psi_{n,m}=\begin{pmatrix}
\psi_{n-1,m}^{(1)} \\
\psi_{n,m}^{(2)} 
\end{pmatrix}$ \\ \hline
{$ \psi_{n,m}^{(1)} =C_1\frac{p^{|m|}\prescript{}{2}F_1\left(-n,n+2+\frac{1}{\lambda\b},|m|+1,\frac{\b p^2}{1+\b p^2}\right)}{\left(1+\b p^2\right)^{1+\frac{1}{2\lambda\b}}}e^{im\vart}$} 
&
{$ \psi_{n,m}^{(2)} =C_2\frac{p^{|m+1|}\prescript{}{2}F_1\left(-n,n+\frac{1}{\lambda\b},|m|,\frac{\b p^2}{1+\b p^2}\right)}{\left(1+\b p^2\right)^{\frac{1}{2\lambda\b}}}e^{i(m+1)\vart}$}  
\end{tabular}
\end{ruledtabular}
\end{table*}
\begin{table*}[t!]
\caption{\label{tab:2negative}Energy levels and the corresponding wave functions for $ m<-\frac{1}{2} -\frac{1}{\lambda\b}$. In this case, similarly to what happens for the levels in Table~\ref{tab:positive} the degeneracy of the energy levels with $n+|m|=N$  and $N\ge [\frac{1}{2}+\frac{1}{\lambda\beta}] + 1$ is finite and given by $D=N-[\frac{1}{2}+\frac{1}{\lambda\beta}]$. } 
\begin{ruledtabular} 
\begin{tabular}{c}
$ E_{n,m}=\sqrt{M^2c^4+2\hbar eB_0 c \left(n+|m|\right)\left[\b\frac{\hbar eB_0}{2c}\left(n+|m|\right)-1\right]}$ \ \ \ $ n=0,1,2,\ldots$\ \ \ 
$\psi_{n,m} =
\begin{pmatrix} 
\psi_{n,m}^{(1)}  \\
\psi_{n,m}^{(2)} 
\end{pmatrix} $\\  \hline 
$\psi_{n,m}^{(1)} =C_1\frac{p^{|m|}\prescript{}{2}F_1\left(-n,n+2|m|-\frac{1}{\lambda\b},1+\left|m\right|,\frac{\b p^2}{1+\b p^2}\right)}{\left(1+\b p^2\right)^{|m|-\frac{1}{2\lambda\b}}}e^{im\vart}$\, \, \, \, \, \,
$\psi_{n,m}^{(2)} =C_2\frac{p^{(|m|-1)}\prescript{}{2}F_1\left(-n,n+2|m|-\frac{1}{\lambda\b},|m|,\frac{\b p^2}{1+\b p^2}\right)}{\left(1+\b p^2\right)^{|m|-\frac{1}{2\lambda\b}}}e^{i(m+1)\vart}$ \\ 
\end{tabular}
\end{ruledtabular}
\end{table*}

\section{spectrum and  wave functions in momentum space}
\label{spectrumWF}

Since we have reduced ourselves to the exact study of a Schr\"{o}dinger equation in the $x$ (or $q$) space, the problem does not need any further inspection. However, for a better understanding as well as for completeness the full spinorial solutions will be given in the $p$-space. 
Starting from tables I and II and the $\varphi^{(i)}$ wave functions in the $q$ space, we simply obtain the form of the corresponding "radial" wave functions in the $p$-space through Eq.~\eqref{ufun} and Eq.~\eqref{rx}. They are found to be of the form ($C_i$ is a normalization constant)
\begin{multline}
\label{wfp}
u^i_{n,m}(p)=C_i\frac{p^{\ze_i-\frac{1}{2}}}{(1+\b p^2)^{\frac{\ze_i+\xi_i}{2}}}\, \times\\ \,\phantom{xxxF}_{2}F_1\left(-n,\ze_i+\xi_i+n,\ze_i+\frac{1}{2};\frac{\b p^2}{1+\b p^2}\right)
\end{multline}
where $\zeta_i$ and $\xi_i$ are defined by Eq.~\eqref{xizeta1} and Eq.~\eqref{xizeta2} and $_2F_1$ is the hypergeometric series~\cite{abramowitz}. 
We shall now classify the eigenfunctions and the corresponding energy values according to the angular quantum number $m$. The results are summarized in tables \ref{tab:positive}, \ref{tab:1negative} and \ref{tab:2negative}. We give here explicitly only the positive branch of the spectrum. For the negative eigenvalues similar formulas are readily obtained (see Eq.~\eqref{k}).  

Then, we build up the full spinor solutions by putting together those states of $\psi^{(1)}$ and $\psi^{(2)}$ which have the same energy eigenvalues. In doing this one has to use the last columns of Tables \ref{tab:up} and \ref{tab:down}. 
It turns out that the first row of Table \ref{tab:up} provides the same energy spectrum as the corresponding row of Table \ref{tab:down}: together, they produce Table \ref{tab:positive}. Analogously for the second and third rows which respectively imply Tables \ref{tab:1negative} and \ref{tab:2negative}. One can verify that the upper and lower spinorial components are actually interlaced by (\ref{PpandPm}).
The ranges for the quantum number $m$ in Tables \ref{tab:positive}, \ref{tab:1negative} and \ref{tab:2negative} descend from the first column of Tables \ref{tab:up} and \ref{tab:down}. When coupling the first row of Table \ref{tab:up} and the first of Table \ref{tab:down}, one takes the intersection set of the two $m$-domains. The same is done for the other rows. This produces the three different possible sets of values of $m$ presented in Tables \ref{tab:positive}, \ref{tab:1negative} and \ref{tab:2negative}.

{\color{black}
Note also that in Eq.~\eqref{wfp} and tables~\ref{tab:positive}, \ref{tab:1negative} and~\ref{tab:2negative} the normalization constants $C_i$, with  $(i=1,2)$, are not independent. This can be seen from the fact that the two spinor components $\psi^{(i)}$ must satisfy the intertwining relations \eqref{PpandPm} or from the fact that requiring a given normalization for the full spinor solution gives a relation between $C_1$ and $C_2$.
}

Let us now examine and discuss the results presented in these Tables.

Table \ref{tab:positive} gives the energies and the eigenfunctions with positive values of the angular momentum quantum number ($m\ge 0$). We find that all the energy levels except the lowest state (which is a singlet) have a finite degeneracy. For example, for $n+m=N$ the levels are $(N+1)$-fold degenerate. Also, all states are doublets i.e. have a spin up as well as a corresponding spin down component. 
Table~\ref{tab:1negative}  shows the results for values of the angular momentum quantum number in the range $-\frac{1}{2}-\frac{1}{\lam\b}<m\leq -1$.  

Note that:
\begin{eqnarray}
\frac{1}{\lam\b}&=&\frac{2c}{\b\hbar e B_0}=\frac{2\hbar Mc}{\left(\hbar\sqrt{\b}\right)^2 eB_0 M}=2\left(\frac{l_c}{\hbar\sqrt{\b}}\right)^2\nonumber \\ &=&2\left(\frac{l_c}{\Delta x_0}\right)^2 \gg 1,~~~~l_c=\sqrt{\frac{\hbar}{M\omega_L}}\label{constr}
\end{eqnarray}
where $\omega_L=\frac{eB_0}{Mc}$ denotes the electron cyclotron frequency and $l_c$ is just the characteristic length of the associated oscillator, which has to be considerably larger than the minimal observable length if this very problem has to be studied. Therefore $m$ can not assume an arbitrarily low negative value but is constrained by the lower limit $(-\f{1}{2}-\f{1}{\lam\b})$ (which is a very large negative number by virtue of (\ref{constr}) and in the limit $\beta \rightarrow 0$ it becomes infinitely negative).
These class of energy levels have a finite degeneracy $D=[\frac{1}{2}+\frac{1}{\lambda \beta}]$ for finite values of $\beta$.  $D$ becomes infinitely large when $\beta \to 0$ and this family of states reduces to the ordinary quantum states of the relativistic Landau problem with negative values of $m$. Interestingly the ground state is a spin singlet while the excited states are spin doublets. 
 Table~\ref{tab:2negative} gives the energy eigenvalues and eigenfunctions with $m$ in the range $ m<-\frac{1}{2} -\frac{1}{\lambda\b}$. We note that this range becomes meaningless when $\beta \to 0$ and the corresponding states loose therefore any physical meaning in this limit.
However for finite values of $\beta$ (a non zero minimal length) such states are physical states and must be included in the physical spectrum. They are all doublet states and the energy levels for which  $n+|m|= N$ with $N\geq [\frac{1}{2} +\frac{1}{\lambda\beta}]+ 1$  the degeneracy is given by: $D=N-[\frac{1}{2}+\frac{1}{\lambda\beta}]$.

 From the above tables it can also be seen that for $m=-\f{1}{2}-\f{1}{\lam\b}$ there isn't any acceptable spinorial solution. This directly descends from our $q$-space analysis where one observes that such a value of $m$ would make it necessary to appeal to the second line in Table \ref{tab:up} (for the upper component) and to the third line in Table \ref{tab:down} (for the lower one). These solutions cannot be coupled though, as it is straightforward to verify that they don't share the same energy, or in other words that the corresponding $p$-space components $\psi^{(1)}$ and $\psi^{(2)}$ thus obtained do not verify (\ref{PpandPm}).
 
We observe here that the no-GUP context is correctly reproduced by letting $\beta\ra 0$, because in this case Table \ref{tab:2negative} along with its angular momentum domain of validity becomes meaningless, and the degeneracy D for the negative-$m$ solutions of Table \ref{tab:1negative} approaches infinity. Note also that in the mentioned limit the following relation holds between the hypergeometric and the confluent hypergeometric series
\beq
\lim_{\b\ra 0}\,\prescript{}{2}F_{1}\left(-n,{\color{black}\kappa}+\frac{1}{\b\lam},\gamma;\frac{\b p^2}{1+\b p^2}\right)=\prescript{}{1}F_{1}\left(-n,\gamma;\frac{p^2}{\lam}\right)
\eeq
as can be straightforwardly checked from their standard definitions
\begin{widetext}
\begin{equation}
\phantom{F}_{2}F_1\left(a,b,c;z\right)=
\sum_{k=0}^{\infty}\frac{(a)_k (b)_k}{(c)_k}\frac{z^k}{k!} \qquad \phantom{F}_{1}F_1\left(a,c;x\right)=\sum_{k=0}^{\infty}\frac{(a)_k}{(c)_k}\frac{z^k}{k!} \qquad \vert z \vert < 1
\end{equation}
\end{widetext} 
so that
\beq
\lim_{\b\ra 0} (  {\color{black}\kappa}   +\frac{1}{\b\lam})_k \left(\frac{\b p^2}{1+\b p^2} \right)^k= \left(\frac{p^2}{\lam}\right)^k
\eeq
and the expected eigenfunctions of the ordinary quantum mechanical treatment are obtained.
\begin{table*}[t]
\caption{\label{tab:massless}Energy levels for massless electrons. The degeneracy of the the energy levels in the massless case is similarly discussed as in Tables~\ref{tab:positive},~\ref{tab:1negative} and~\ref{tab:2negative} .}
\begin{ruledtabular}
\begin{tabular}{c l l} 
$ m\geq 0$ & $ n=0,1,2,\ldots$ & $ E_{n,m}=v_F\sqrt{\frac{2\hbar eB_0}{c}\left(n+m\right)\left[1+\b\frac{\hbar eB_0}{2c}\left(n+m\right)\right]}$ \\
\hline
\multirow{2}{3cm}{$-\frac{1}{2}-\frac{1}{\lam\b}<m\le-1$} & $n=0$ & $E_{0}=0$  \\ \cline{2-3} 
	&$n=1,2,\ldots$ & $ E_{n}=v_F\sqrt{\frac{2\hbar eB_0}{c}n\left(1+\b\frac{\hbar eB_0}{2c}n\right)}$ \\ \hline
$ m<-\frac{1}{2} -\frac{1}{\lambda\b}$ & $ n=0,1,2,\ldots$ & $ E_{n,m}=v_F\sqrt{\frac{2\hbar eB_0}{c}\left(n+|m|\right)\left[\b\frac{\hbar eB_0}{2c}\left(n+|m|\right)-1\right]}$ \\ 
\end{tabular}
\end{ruledtabular}
\end{table*}

\subsection{Massless Dirac equation in $(2+1)$ dimensions}
\label{graphene}
The equation governing the motion of electrons in graphene is similar to the $(2+1)$ dimensional massless Dirac equation except that the electrons move with Fermi velocity $v_F\approx \frac{1}{300} c$ rather than with the velocity of light $c$. The Hamiltonian for the electrons in graphene in the presence of a magnetic field is
\beq
H_D= v_F \bm{\sigma}\cdot( \hat{\bm p}+\frac{e}{c} \hat{\bm A})
\eeq
In order to obtain the spectrum and the wave functions one needs to make minor changes in the results obtained earlier. The energy levels are presented in Table~\ref{tab:massless} (we only write down the spectra since all the rest stands unchanged).

From Table~\ref{tab:massless} we find that for $m\ge0,n+m=N$ the levels are $(N+1)$ fold degenerate. For $-\frac{1}{2}-\f{1}{\lambda\beta}<m\le-1$ there is a  zero energy ground state which is a spin singlet while  the excited states are spin doublets. All these states have a (large) finite degeneracy with respect to $m$ which is as before given by $D=[\frac{1}{2}+\f{1}{\lambda\beta}]$. Finally, for $ m<-\frac{1}{2} -\frac{1}{\lam\b}$ the degeneracy of the levels with $n+|m|=N$,   $N>\frac{1}{2} +\frac{1}{\lam\b}$ is similarly calculated to be given by:
$D=N-[\frac{1}{2}+\f{1}{\lambda\beta}]$. As in the massive case these solutions become meaningless in the limit of a vanishing minimal length ($\beta \to 0$). {\color{black} Finally to see how the energy levels deviate from the usual relativistic Landau levels, let us examine the spectrum in the small $\b$ limit. For this let us consider the second energy level in Table VI and expanding with respect to $\b$ we find
\begin{equation}
\label{theoryLL}
E_n=v_F\sqrt{\f{2\hbar eB_0}{c}n}+\b v_F\f{\hbar eB_0n}{2c}\sqrt{\f{\hbar eB_0n}{2c}}+{\cal O(\b^2)}
\end{equation}
 where the first term gives the usual Landau levels. Thus the spectrum contains an additional term involving $n^{{3}/{2}}$ and the dispersion relation is indeed modified. {\color{black} These modifications of the dispersion relation might have important implications,  for instance in the calculation of quantities like the density of states, which will be addressed elsewhere.}} 
{\color{black}
Here we would like instead to comment on  how our exact solution of the GUP Dirac equation in 2+1 dimensions in a constant magnetic field might  be already of use in deriving an upper bound on the minimal length.

In order to do so we can compare our derived formulae for the energy spectrum  with the experimental measurement of the transitions between graphene Landau Levels (LL)~\cite{expLL}.

In ref.~\cite{expLL}  infrared spectroscopy has been used to resolve the transitions between graphene LL in the presence of magnetic fields of intensities up to $B_0= 18$ Tesla. The authors report a linear behavior of these transmission resonances with $\sqrt{B_0}$  from which a best fit value of the fermi velocity $v_F= (1.12\pm 0.02)\times 10^6$ m/s is deduced.  From this \emph{experimental} value of the Fermi velocity one can deduce for instance  for the first excited level of the graphene Landau spectrum in the absence of a minimal length (c.f. first term of  Eq.~\eqref{theoryLL} with n=0), $E=v_F\sqrt{\frac{2\hbar eB_0}{c}}$ for $B_0=18 \, \text{T}$ the (experimental) value:
 \begin{equation}
 E=(172\pm3)\, \text{meV}
 \label{experror}
 \end{equation}
 and note that $(\delta E) / E = (\delta v_F) / v_F$.
We wish to use this result to provide an  upper  bound  to the  observable minimal length $\Delta x_0=\hbar\sqrt{\b}$ of our GUP model.  We may use the results of Table VI. Picking the spectrum on the first line and setting $m=0$, $n=1$ one has for the energy of the first graphene LL: 
\begin{eqnarray}
E_{1,0}^{(\beta)}&=&v_F\sqrt{\frac{2\hbar eB_0}{c}\left(1+\beta\frac{\hbar eB_0}{2c}\right)}\nonumber \\ &=& E_{1,0}^{(\beta=0)}\, \sqrt{\left(1+\beta\frac{\hbar eB_0}{2c}\right)}
\end{eqnarray}
The impossibility to experimentally distinguish the deviation brought about  by the existence of a minimal length means that the two values,  predicted ($E_{1,0}^{(\beta)}$) and  experimental  ($E_{1,0}^{(\beta=0)} \sim E$) must be close enough, i.e. they must be, with respect to each other, within the experimental error (c.f. Eq.~\eqref{experror}) hence we can surely assume that:
\begin{equation}\Delta E= E_{1,0}^{(\beta)}-E_{1,0}^{(\beta=0)}< 6 \, \text{meV}\end{equation}
from which:
\begin{eqnarray}
\Delta E &=& E_{1,0}^{(\beta=0)}\left(\sqrt{1+\beta\frac{\hbar eB_0}{2c}}-1\right)\nonumber \\&=&E\left(\sqrt{1+\delta}-1\right)<6\, \,\text{meV} 
\end{eqnarray}
where we have defined $\delta=\beta\frac{\hbar eB_0}{2c}=(\hbar\sqrt{\beta})^2\,\frac{eB_0}{2\hbar c}$. Hence  since  $\delta$ is expected to be a very small quantity we obtain the constraint:
\begin{equation}
\label{rel_error}
\delta < \frac{12}{172} \approx 0.07 
\end{equation}
which in turn, resorting to gaussian units, leads us  (with $B_0= 18$ T = $18 \times 10^{4}$ Gauss) to: 
\begin{equation}
\label{MLbound}
\Delta x_0 =\hbar\sqrt{\b}< 2.3 \,\text{nm}
\end{equation}
thus providing in principle an upper bound on the minimal  length (or equivalently on the parameter $\b$) appearing in the  framework of a generalized uncertainty principle. 

As a final remark we wish to point out that in ref.~\cite{expLL} the authors find some discrepancies on the value of the Fermi velocity deduced from different LL transitions and warn about possible difficulties of the  simple interpretation of IR data in terms of a simple LL energy subtraction based on 
standard one particle quantum mechanical results and conclude that many particles effects may be expected to contribute to the LL transition energies. 
These many particle effects may therefore also affect  the upper bound derived here (c.f. Eq.~\eqref{MLbound}) on the minimal length.

Admittedly the upper bound in Eq.~\eqref{MLbound}  is not a very strong bound.  It is however comparable with those derived in~\cite{Frassino:2011aa} where the upper bound obtained from the Casimir effect for the minimal distance of the plates of 0.5 $\mu$m, depending on the particular GUP model, is in the range $\approx 29-58$ nm.
Similar order of magnitude upper bounds on the scale of phase-space non commutativity have been derived in~\cite{Bastos:2012kh}. 

We could perhaps note that a possibility to make our bound more stringent would be to follow the approach of ref.~\cite{iontrapping} and assume that in future experiments it will be possible to measure LL transitions for very large values of the quantum number $n$.  Our argument that led to Eq.~\eqref{MLbound} could be reproduced for the $n$-th LL level and would provide the bound:
\begin{equation}
\hbar\sqrt{\beta} < \sqrt{\frac{2}{n} \,  \frac{(\Delta E_n)_\text{exp}}{(E_n)_{\text{exp}}}\, \frac{2 c}{e B_0}} = \frac{2.3\ \text{nm}}{\sqrt{n}}
\end{equation}
(assuming, somewhat optimistically, the same value for the relative error of the measure of $E_n$ as for the first excited level, c.f.~\eqref{rel_error}). 
As discussed in Ref.~\cite{BRAU} a rather constraining upper
bound on the minimal length comes also from the hydrogen $1S-2S$ transition: $(\Delta x)_{min}=\hbar\sqrt{\beta} < 10^{-2}\ $fm =$10^{-17}\ $m.  The authors of~\cite{BRAU} argue that this bound could be avoided by assuming that the parameter $\beta $ is not a universal constant and could vary
from one system to another depending, for example, on the
energy content of the system (the mass of the particle, for
instance) or the strength of some interaction. Indeed, in
\cite{BRAU}, by making this hypothesis, the authors, through a
comparison with the experimental results for ultracold
neutron energy levels in a gravitational quantum well
(GRANIT experiment)~\cite{GRANIT}, derive a relaxed upper
bound to the minimal length which turns out to be of the
order of a few nanometers [$(\Delta x)_{min} < 2.41$ nm], which is quite close to the one derived here (c.f.~\eqref{MLbound}). Clearly we could as well advocate the non universality of $\beta$  in order to evade the stronger constraints as those discussed in~\cite{BRAU} and also in~\cite{Benczik:2005bh,Stetsko:2006mm}.
}
\section{Discussion and Conclusion}
\label{discussion}

We have obtained exact solutions of the $(2+1)$ dimensional Dirac equation in an external homogeneous magnetic field in the presence of a minimal length. 
We work within a momentum space representation of the Heisenberg algebra and through an appropriate transformation of both the wave function and the variable the  second order equations for the Dirac components are reduced into a finite domain Schr\"odinger like exactly solvable problem (trigonometric Scarf potential).  Interestingly it is shown that the ordinary boundary conditions in the finite domain (vanishing of the wave function at the end-points) can be transported back to the radial $p$-space and interpreted in terms of the finiteness of the energy integral.

The solutions show that a non-zero  minimal length changes the spectrum to a large extent as compared to the standard relativistic Landau problem. A notable feature of this problem is that when the angular momentum quantum number $m$ is negative it is constrained and different ranges of its value point to different class of physical states.  However, the constraint on the quantum number $m$ disappears as the minimal length vanishes ($\beta \rightarrow 0$). This can be seen in Tables \ref{tab:1negative} and \ref{tab:2negative}. Another feature worth noting is the degeneracy pattern of the energy levels. In the usual relativistic Landau problem, the Landau levels are infinitely degenerate for $m < 0$. In contrast, in the present case  some energy levels are finitely degenerate (as in Table \ref{tab:positive}) while others have very large finite degeneracy (Table \ref{tab:1negative}). It may also be noted that an interesting feature of the minimal length scenario turns out to be the appearance of the solutions reported in Table \ref{tab:2negative}. Indeed these solutions exist only for $\b\neq 0$. In the limit $\b\ra 0$ the related range of $m$ becomes meaningless and also the corresponding eigenfunctions are no longer physically acceptable. In this limit, the correct non minimal length situation can be recovered from the results  of Tables \ref{tab:positive} and \ref{tab:1negative}. 

{\color{black} We have briefly discussed how our exact solution of the problem in the massless case might be used to provide an upper bound on the minimal length via a comparison with exisiting experimental measurements of transitions between graphene LL. }
Finally we wish to point out that, always in the massless case, it would be of interest to compute {\color{black}other} physical quantities e.g, Hall conductivity where a comparison with experimental results may provide {\color{black}perhaps more stringent} bounds on the minimal length $\hbar\sqrt{\beta}$ {\color{black}than those discussed here}. 
\begin{widetext}
\begin{acknowledgments}
This work is an outcome of the diploma thesis of L.~M. presented at the University of Perugia in September 2012. P.~R. acknowledges hospitality from the Physics Department of the
University of Perugia and INFN - Istituto Nazionale di Fisica Nucleare - Sezione di Perugia for financial support.
\end{acknowledgments}
\end{widetext}

\ed